\documentclass[aps,prl,twocolumn,superscriptaddress,nofootinbib]{revtex4}

\usepackage{graphicx}
\usepackage{mathptmx, textcomp}
\usepackage[usenames]{color}
\usepackage{graphicx}  
\usepackage{dcolumn}   
\usepackage{bm}        
\usepackage{amssymb, amsmath}   
\usepackage[usenames]{color}
\usepackage[colorlinks,urlcolor=blue,citecolor=blue,linkcolor=blue]{hyperref}

\begin{document}

\title{Metastability and Coherence of Repulsive Polarons in a Strongly Interacting Fermi Mixture}

\author{C. Kohstall}
\affiliation{Institut f\"ur Quantenoptik und Quanteninformation (IQOQI), \"Osterreichische Akademie der Wissenschaften, 6020 Innsbruck, Austria}
\affiliation{Institut f\"ur Experimentalphysik und Zentrum f\"ur Quantenphysik, Universit\"at Innsbruck, 6020 Innsbruck, Austria}

\author{M. Zaccanti}
\affiliation{Institut f\"ur Quantenoptik und Quanteninformation (IQOQI), \"Osterreichische Akademie der Wissenschaften, 6020 Innsbruck, Austria}

\author{M. Jag}
\affiliation{Institut f\"ur Quantenoptik und Quanteninformation (IQOQI), \"Osterreichische Akademie der Wissenschaften, 6020 Innsbruck, Austria}

\author{A. Trenkwalder}
\affiliation{Institut f\"ur Quantenoptik und Quanteninformation (IQOQI), \"Osterreichische Akademie der Wissenschaften, 6020 Innsbruck, Austria}

\author{P. Massignan}
\affiliation{ICFO - Institut de Ci\`encies Fot\`oniques, Mediterranean Technology Park, 08860 Castelldefels (Barcelona), Spain}

\author{G. M. Bruun}
\affiliation{Department of Physics and Astronomy, University of Aarhus, 8000 Aarhus C, Denmark}

\author{F. Schreck}
\affiliation{Institut f\"ur Quantenoptik und Quanteninformation (IQOQI), \"Osterreichische Akademie der Wissenschaften, 6020 Innsbruck, Austria}

\author{R. Grimm}
\affiliation{Institut f\"ur Quantenoptik und Quanteninformation (IQOQI), \"Osterreichische Akademie der Wissenschaften, 6020 Innsbruck, Austria}
\affiliation{Institut f\"ur Experimentalphysik und Zentrum f\"ur Quantenphysik, Universit\"at Innsbruck, 6020 Innsbruck, Austria}

\newcommand{\bra}[1]{\mbox{\ensuremath{\langle #1 \vert}}}
\newcommand{\ket}[1]{\mbox{\ensuremath{\vert #1 \rangle}}}
\newcommand{\mb}[1]{\mathbf{#1}}
\newcommand{\phipp}{\big|\phi_{\mb{p}}^{(+)}\big>}
\newcommand{\phipav}{\big|\phi_{\mb{p}}^{\p{av}}\big>}
\newcommand{\pp}[1]{\big|\psi_{p}(#1)\big>}
\newcommand{\drdy}[1]{\sqrt{-R'(#1)}}
\newcommand{\Rb}{$^{87}$Rb}
\newcommand{\K}{$^{40}$K }
\newcommand{\Li}{$^{6}$Li }
\newcommand{\LiK}{$^6$Li-$^{40}$K}
\newcommand{\na}{${^{23}}$Na}
\newcommand{\muK}{\:\mu\textrm{K}}
\newcommand{\p}[1]{\textrm{#1}}
\newcommand\T{\rule{0pt}{2.6ex}}
\newcommand\B{\rule[-1.2ex]{0pt}{0pt}}
\newcommand{\reffig}[1]{\mbox{Fig.~\ref{#1}}}
\newcommand{\refeq}[1]{\mbox{Eq.~(\ref{#1})}}
\hyphenation{Fesh-bach}
\newcommand{\previous}[1]{}
\newcommand{\note}[1]{\textcolor{red}{[\textrm{#1}]}}
\newcommand{\gbox}{\textcolor{ForestGreen}{$\Box$}}
\newcommand{\bcirc}{\textcolor{blue}{$\bigcirc$}}
\newcommand{\cit}{\textcolor{red}{[cite]}}

\date{\today}



\maketitle

%


{\bf
Ultracold Fermi gases with tuneable interactions represent a unique test bed to explore the many-body physics of strongly interacting quantum systems \cite{Bloch2008mbp, Giorgini2008tou, Radzihovsky2010ifr, Chevy2010ucp}. In the past decade, experiments have investigated a wealth of intriguing phenomena, and precise measurements of ground-state properties have provided exquisite benchmarks for the development of elaborate theoretical descriptions. Metastable states in Fermi gases with strong repulsive interactions \cite{Duine2005ifi, Leblanc2009rfg, Conduit2009ipf, Jo2009ifi, Pilati2010ifo, Chang2011fit, Sanner2011cap} represent an exciting new frontier in the field. The realization of such systems constitutes a major challenge since a strong repulsive interaction in an atomic quantum gas implies the existence of a weakly bound molecular state, which makes the system intrinsically unstable against decay.
Here, we exploit radio-frequency spectroscopy to measure the complete excitation spectrum of fermionic \K impurities resonantly interacting with a Fermi sea of \Li atoms. In particular, we show that a well-defined quasiparticle exists for strongly repulsive interactions. For this ``repulsive polaron'' \cite{Pilati2010ifo, Massignan2011rpa, Schmidt2011esa} we measure its energy and its lifetime against decay. We also probe its coherence properties by measuring the quasiparticle residue. The results are well described by a theoretical approach that takes into account the finite effective range of the interaction in our system. We find that a non-zero range of the order of the interparticle spacing results in a substantial lifetime increase.  This major benefit for the stability of the repulsive branch opens up new perspectives for investigating novel phenomena in metastable, repulsively interacting fermion systems.
}

Landau's theory of a Fermi liquid \cite{Landau1957tto} and the underlying concept of quasiparticles lay at the heart of our understanding of interacting Fermi systems over a wide range of energy scales, including liquid $^{3}$He, electrons in metals, atomic nuclei, and the quark-gluon plasma.
In the field of ultracold Fermi gases, the normal (non-superfluid) phase of a strongly interacting system can be interpreted in terms of a Fermi liquid \cite{Lobo2006nso, Schirotzek2009oof, Navon2010teo, Nascimbene2011flb}. In the population-imbalanced case, quasiparticles coined Fermi polarons are the essential building blocks and have been studied in detail experimentally \cite{Schirotzek2009oof} for attractive interactions. Recent theoretical work \cite{Pilati2010ifo, Massignan2011rpa, Schmidt2011esa} has suggested a novel quasiparticle associated with {\em repulsive} interactions. The properties of this repulsive polaron are of fundamental importance for the prospects of repulsive many-body states. A crucial question for the feasibility of future experiments is the stability against decay into molecular excitations \cite{Pekker2011cbp, Massignan2011rpa, Sanner2011cap}. Indeed, whenever a strongly repulsive interaction is realized by means of a Feshbach resonance \cite{Chin2010fri}, a weakly bound molecular state is present into which the system may rapidly decay.

Our system consists of impurities of fermionic \K atoms immersed in a large Fermi sea of $^6$Li atoms, which is characterised by a Fermi energy $\epsilon_F = h \times 37\,$kHz and a temperature $T = 0.16\, \epsilon_F/k_B$ (see Methods), with $h$ and $k_B$ denoting Planck's and Boltzmann's constants. In a particular combination of spin states \cite{Naik2011fri}, the \Li-\K mixture features a Feshbach resonance centered at $B_0 = 154.719(2)$\,G. The resonance allows to widely tune the $s$-wave interaction, parametrised by the scattering length $a$, via a magnetic field $B$. The interaction strength is described by the dimensionless parameter $-1/(\kappa_F a)$, where $\kappa_F = \hbar^{-1} \sqrt{2 m_{\rm Li} \epsilon_F} = 1/(2850\,a_0)$ 
is the Fermi wave number; here $\hbar = h/2\pi$, $a_0$ is Bohr's radius, and $m_{\rm Li}$ is the mass of a $^6$Li atom. Near the resonance center, the linear approximation $-1/(\kappa_F a) = (B-B_0)/20\,{\rm mG}$ holds. The momentum dependence of the interaction is characterised by the effective range, which we express in terms of the parameter \cite{Petrov2004tbp} $R^* = 2700\,a_0$ (see Supplementary Information).

\begin{figure}
\vskip 0 pt
\includegraphics[width=8.5cm]{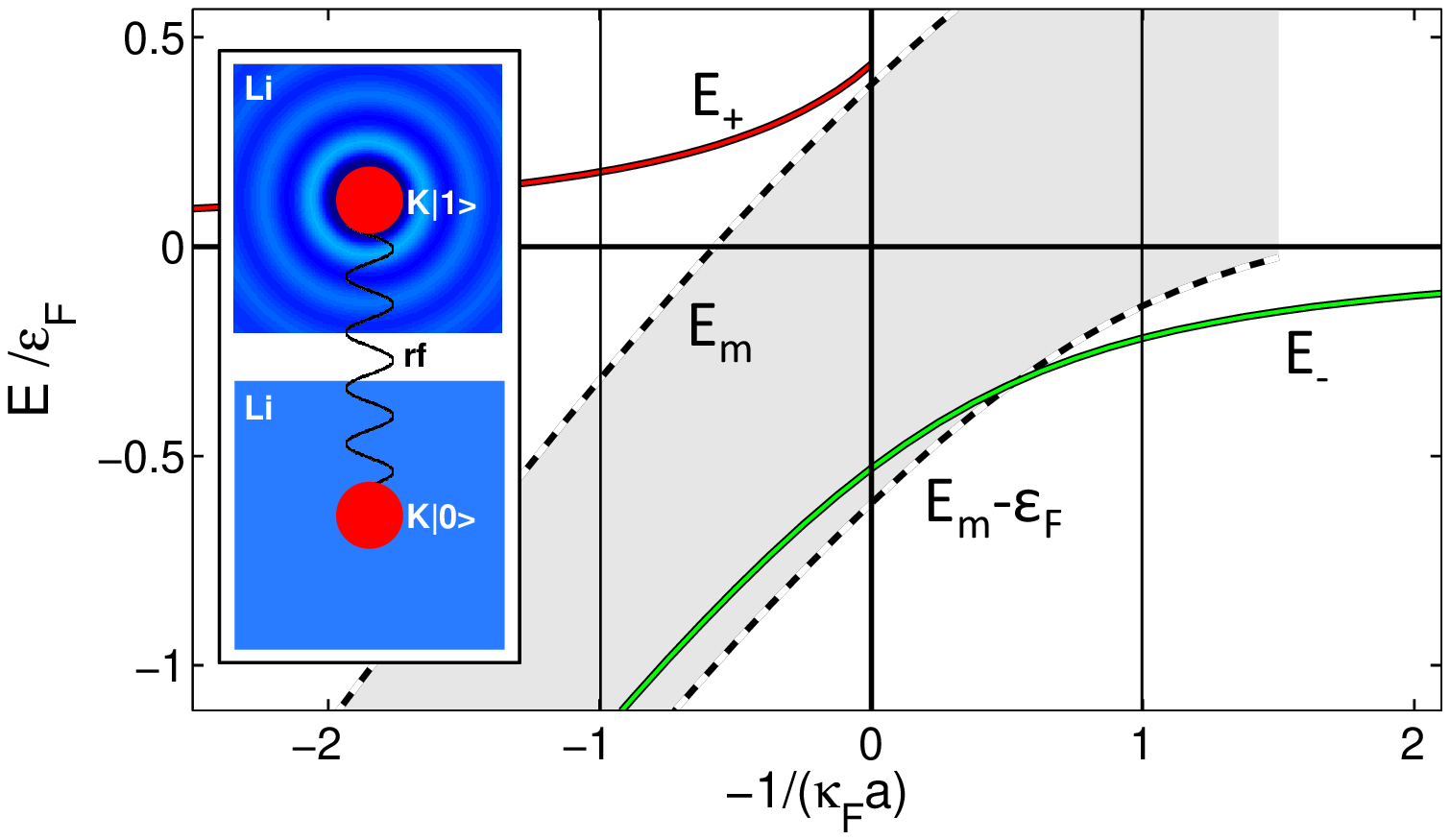}
\caption{\textbf{Energy spectrum of the impurity in the Fermi sea.} For the two polaronic branches, the energies $E_+$ (red line) and $E_-$ (green line) are plotted as a function of the interaction parameter $-1/(\kappa_F a)$. The shaded area between the dashed lines representing $E_m$ and $E_m - \epsilon_F$ (see text) shows the continuum of molecular excitations. The vertical lines at $1/(\kappa_F a) = \pm 1$ indicate the width of the strongly interacting regime. The inset illustrates our rf spectroscopic scheme where the impurity is transferred from a noninteracting spin state $|0\rangle$ to the interacting state $|1\rangle$.}
\label{Figure1}
\end{figure}

Figure~1 illustrates the basic physics of our impurity problem in the $T=0$ limit, showing the energies of different states as a function of the interaction parameter. The situation is generic for any impurity in a Fermi sea, but quantitative details depend on both the mass ratio and the particular width of the Feshbach resonance. The theoretical curves are based on an extension of an approach presented in Refs.~\cite{Punk2009ptm,Massignan2011rpa} to our case of a relatively narrow Feshbach resonance with $\kappa_F R^* = 0.95$ and thus a considerable effective range of the interaction (see Supplementary Information).

The spectrum exhibits two quasiparticle branches, which do not adiabatically connect when the resonance is crossed, and a molecule-hole continuum (MHC).
The interaction-induced energy shifts of the two branches ($E_+ >0$ and $E_- <0$) are generally described in a many-body picture by dressing the impurities with particle-hole excitations. Far away from resonance this simplifies to a mean-field shift proportional to $a$. The lower branch $E_-$ of the system (green line) corresponds to the {\em attractive polaron}, which has recently received a great deal of attention theoretically \cite{Chevy2010ucp, Lobo2006nso, Combescot2007nso, Punk2009ptm, Sadeghzadeh2011mis} as well as experimentally \cite{Nascimbene2009coo, Schirotzek2009oof, Navon2010teo}. This polaronic branch remains the ground state of the system until a critical interaction strength is reached, where the system energetically prefers to form a bosonic molecule by binding the impurity to an atom taken from the Fermi sea \cite{Prokofev2008fpp, Punk2009ptm, Sadeghzadeh2011mis}. The continuum arises from the fact that a majority atom with an energy between $0$ and $\epsilon_F$ can be removed from the Fermi sea to form the molecule. The MHC thus exists in an energy range between $E_m$ and $E_m - \epsilon_F$ (dashed lines in Fig.~1), where $E_m$ represents the energy of a dressed molecule including the binding energy of a bare molecule in vacuum and a positive interaction shift. The attractive polaron can decay into a molecular excitation if this channel opens up energetically ($E_- \ge E_m - \epsilon_F$).

The upper branch (red line in Fig.~1) corresponds to the \emph{repulsive polaron} \cite{Pilati2010ifo, Massignan2011rpa, Schmidt2011esa} with an energy $E_+ >0$. Approaching the resonance from the $a>0$ side, $E_+$ gradually increases and reaches a sizeable fraction of $\epsilon_F$. However, the polaronic state becomes increasingly unstable as it decays to the lower lying states (attractive polaron and MHC). Close to the resonance center, the polaronic state becomes ill-defined as the decay rate approaches $E_+/\hbar$.

To investigate the excitation spectrum of the impurities, we employ radio-frequency (rf) spectroscopy \cite{Chin2004oop, Shin2007trf, Stewart2008ups}. We initially prepare the \K atoms in a non-interacting spin state $|0\rangle$ and then, with a variable frequency $\nu_{\rm rf}$, drive rf transitions into the resonantly interacting state $| 1 \rangle$. Our signal is the fraction of atoms transferred, measured as a function of the rf detuning $\nu_{\rm rf} -\nu_0$ with respect to the unperturbed transition frequency $\nu_0$ between the two spin states.
This excitation scheme provides access to the full energy spectrum of the system.
In particular, it allows us to probe the metastable repulsive polaron as well as all states in the MHC. We furthermore take advantage of the {\em coherence} of the excitation process by driving Rabi oscillations. As an important practical advantage, this enables very fast and efficient transfer of population into a short-lived quasiparticle state by application of $\pi$-pulses. Moreover, we will show that measurements of the Rabi frequency directly reveal quasiparticle properties.

\begin{figure}
\begin{center}
\includegraphics[width=8.5cm]{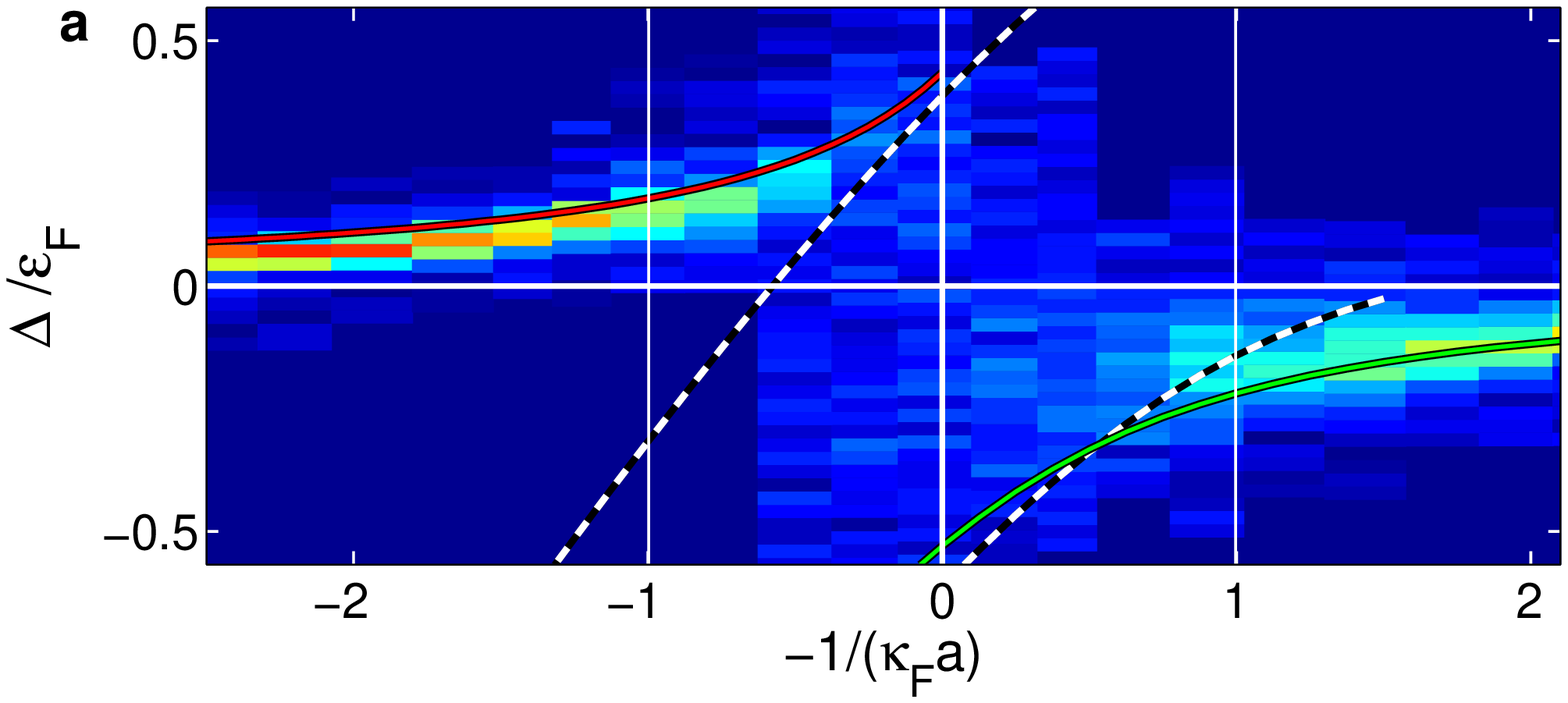}
\includegraphics[width=8.5cm]{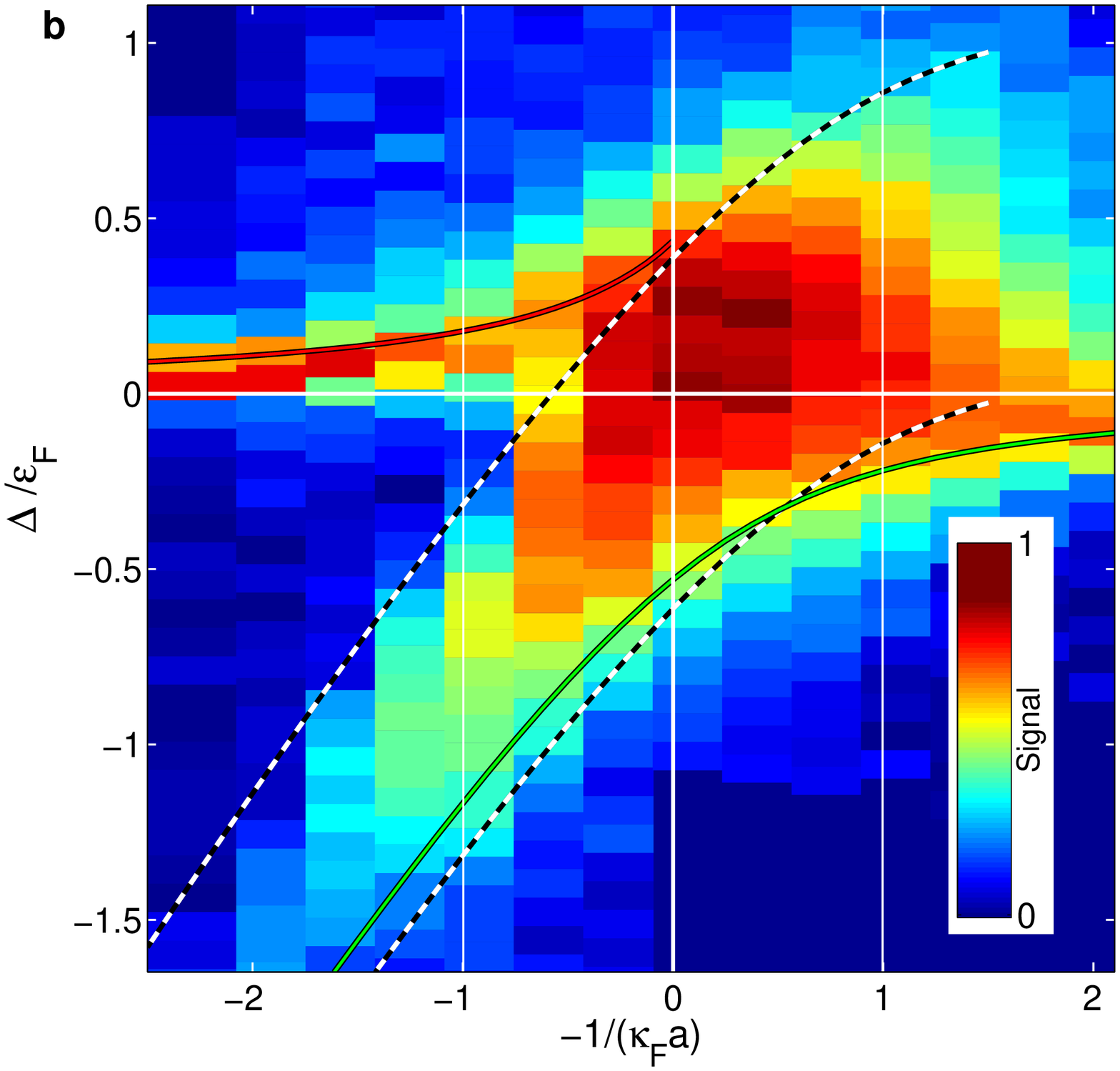}
\caption{\textbf{Spectral response of \K impurities in a \Li Fermi sea.} The false-colour plots show the fraction of \K atoms transferred from the non-interacting spin state $|0\rangle$ into the interacting state $|1\rangle$ for different values of the rf detuning parameter $\Delta = h (\nu_{\rm rf} -\nu_0)$ and for variable interaction strength $-1/(\kappa_F a)$. The panels \textbf{a} and \textbf{b} refer to low and high rf power. For comparison, the lines correspond to the theoretical predictions for $E_+$, $E_-$, $E_m$, and $E_m-\epsilon_F$ as shown in Fig.~1.}
\label{fig2}
\end{center}
\end{figure}

In Fig.~2 we show false-colour plots of our signal, detected for different values of the detuning parameter $\Delta = h (\nu_{\rm rf} -\nu_0)$ and for variable interaction strength $-1/(\kappa_F a)$. Figure 2a displays a set of measurements that was optimised for signal and spectral resolution of the polaronic excitations by using moderate rf power (see Methods). The spectrum in Fig.~2b was optimised for detection of the molecular excitations. Here a much higher rf power had to be employed because of the reduced Franck-Condon wavefunction overlap. For the polaronic branches the high rf power leads to a highly nonlinear saturation behaviour.

Our data clearly show {\em both} polaronic branches, with their measured energies being in excellent agreement with theory. The attractive polaron is found to disappear in the strongly interacting regime. This behaviour, which is different from the one observed for \Li spin mixtures \cite{Schirotzek2009oof}, is consistent with the crossing of $E_-$ and $E_m-\epsilon_F$ at $-1/(\kappa_F a) \approx +0.6$ predicted for our system. In contrast, the repulsive polaron extends far into the strongly interacting regime. A sharp peak is observed in the spectrum with decreasing signal strength until it finally fades out very close to resonance at $-1/(\kappa_F a) \simeq -0.3$ (see Supplementary Information). The low rf power only produces a weak signal for the MHC, whereas the high rf power clearly unveils the presence of the MHC. Further on the $a>0$ side of the resonance, the molecular signal gets weaker because of the decreasing Franck-Condon overlap, and outside of the strongly interacting regime the situation corresponds to the rf association of bare molecules (see Supplementary Information).

\begin{figure}
\begin{center}
\includegraphics[width=8.5cm]{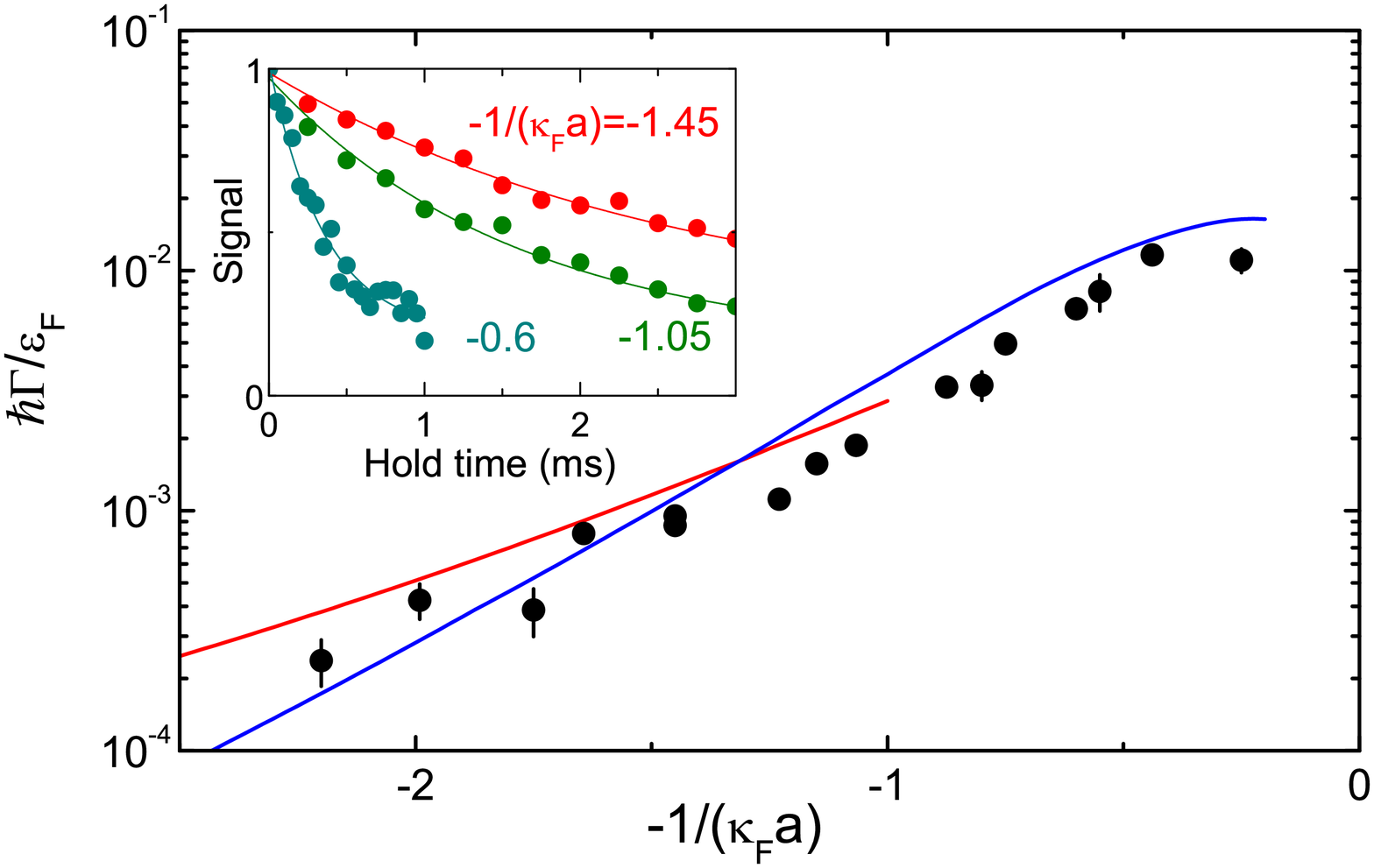}
\caption{\textbf{Decay rate of the repulsive polaron.} The data points display the measured decay rates $\Gamma$ as extracted by exponential fits from decay curves; the error bars indicate the fit uncertainties. Sample decay curves are shown in the inset. The solid lines represent theoretical calculations of the two-body decay (blue line) and the three-body decay (red line) into the attractive polaron or the MHC, respectively.}
\label{fig3}
\end{center}
\end{figure}

To investigate the decay of the repulsive branch, we apply an rf pulse sequence to selectively convert repulsive polarons back into non-interacting impurities after a variable hold time (see Methods). The back conversion sensitively depends on the rf resonance condition and thus allows us to discriminate \K atoms in the polaronic state against those ones forming molecules. Figure~3 presents the experimental results. The inset shows three sample curves, taken for different values of the interaction parameter. The main panel displays the values extracted for the decay rate $\Gamma$ from the decay curves by simple exponential fits. The data reveal a pronounced increase of decay as the resonance is approached, which is in good agreement with theoretical model calculations \cite{Massignan2011rpa} (see Supplementary Information). The decay populates the MHC and may happen in a two-step process where the repulsive polaron first decays via a two-body process into an attractive polaron (blue line) and then decays into a molecular excitation. Alternatively, a three-body process may directly lead into the MHC (red line). Very close to the resonance, for $-1/(\kappa_F a) = -0.25$, we find $\hbar \Gamma / \epsilon_F \approx 0.01$, which corresponds to a $1/e$ lifetime of about $400\,\mu$s. Relating this decay rate to the corresponding energy shift $E_+=0.30\,\epsilon_F$, we obtain $\hbar \Gamma/E_+ \approx 0.03 \ll 1$, which demonstrates that the repulsive polaron exists as a well resolved, metastable quasiparticle even deep in the strongly interacting regime.

\begin{figure}
\begin{center}
\includegraphics[width=7cm]{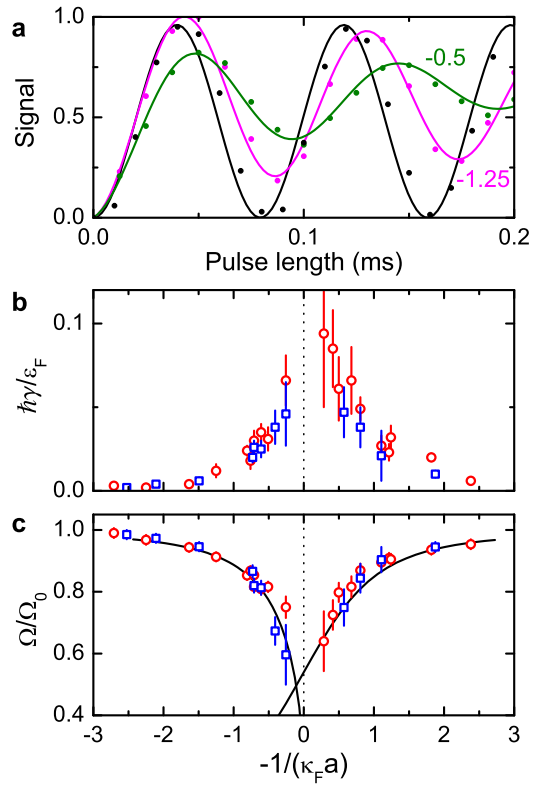}
\caption{\textbf{Rabi oscillations and the quasiparticle residue.} \textbf{a}, Sample Rabi oscillations (magenta and green data points for $-1/(\kappa_F a) = -1.25$ and $-0.5$, respectively) with harmonic oscillator fits (solid lines) demonstrate the two effects of the interaction with the Fermi sea: damping and a reduction of the Rabi frequency. The black curve is a reference curve recorded without \Li. In \textbf{b} and \textbf{c}, the data points show the damping rates $\gamma$ and the normalized Rabi frequencies $\Omega/\Omega_0$ as measured for two different values of the rf power; the blue squares and red dots refer to $\Omega_0 = 2\pi \times 6.5$\,kHz and $12.6$\,kHz, respectively.
The error bars indicate the fit uncertainties. The solid lines represent the theoretical behavior of $\sqrt{Z}$ for the repulsive and the attractive polaron.}
\label{fig4}
\end{center}
\end{figure}

The lifetime observed for the repulsive branch appears to be remarkably long, when compared to recent experiments on \Li spin mixtures \cite{Sanner2011cap}. The latter system is a mass-balanced one and it features a broad Feshbach resonance with a negligible effective range. Our theoretical approach allows to answer in a general way the question how mass imbalance and the width of the resonance influence the lifetime. We find that, while the mass imbalance does only play a minor role \cite{Massignan2011rpa}, the dominant effect results from the finite effective range, which is associated with the narrow character of the Feshbach resonance that we exploit. Comparing our situation with a hypothetical system with a broad Feshbach resonance, and thus with a zero-range interaction, we find that in the strongly interacting regime the same amount of energy can be obtained with an almost ten times increased lifetime (see Supplementary Information).

Besides energy and lifetime, the polaron is characterized by its effective mass $m^*$ and its quasiparticle residue $Z$. The difference between effective and bare mass \cite{Massignan2011rpa} does not produce any significant features in our rf spectra. The residue $Z$ ($0 \le Z \le 1$) quantifies how much of the non-interacting particle is contained in the polaron's wavefunction, which can be written as $\sqrt{Z}\,|1\rangle$ plus terms describing excitations in the Fermi sea. The pre-factor $\sqrt{Z}$ directly manifests itself in the Rabi frequency $\Omega$ that describes the coherent rf coupling between the noninteracting and the polaronic state (see Supplementary Information).

Figure 4 presents the experimental data on Rabi oscillations for variable interaction strength. The sample curves in Fig.~4a demonstrate both the interaction-induced change in the frequency and a damping effect. We apply a simple harmonic oscillator model (including a small increasing background) to analyse the curves, which yields the damping rate $\gamma$ and the frequency $\Omega$. The damping strongly increases close to the resonance center, but does not show any significant dependence on $\Omega_0$, see Fig.~4b. It is interesting to note that the population decay rates $\Gamma$ measured for the repulsive branch (Fig.~3) stay well below the values of $\gamma$, which points to collision-induced decoherence as the main damping mechanism.

Figure 4c displays the measured values for the Rabi frequency $\Omega$, normalized to the unperturbed value $\Omega_0$. The interaction-induced reduction of $\Omega/\Omega_0$ is found to be independent of the particular value of $\Omega_0$ (comparison of blue squares and red dots; see also Supplementary Information). The solid lines show $\sqrt{Z}$ as calculated within our theoretical approach for both the repulsive and the attractive polaron. The comparison with the experimental data demonstrates a remarkable agreement with the relation $\sqrt{Z} = \Omega/\Omega_0$. Our results therefore suggest measurements of the Rabi frequency as a precise and robust method to determine the quasiparticle residue $Z$, and thus provides a powerful alternative to methods based on the detection of the narrow quasiparticle peak in the spectral response \cite{Schirotzek2009oof, Punk2009ptm}.

In conclusion, we have realized an ultracold model system of \K and \Li atoms to investigate the quasiparticle behavior of heavy impurities resonantly interacting with a Fermi sea of light particles. Our spectroscopic approach has confirmed the existence of the predicted repulsive branch \cite{Pilati2010ifo, Massignan2011rpa, Schmidt2011esa} and has demonstrated that the repulsive polaron can exist as a well-defined quasiparticle even deep in the strongly interacting regime. The long lifetime of the repulsive polaron in our system, which we ascribe to the finite effective range of the interparticle interaction, may be a key factor to overcome the problem of decay into molecular excitations \cite{Pekker2011cbp, Sanner2011cap} in the experimental investigation of metastable many-body states that rely on repulsive interactions. In particular, the creation of states with two fermionic components phase-separating on microscopic or macroscopic scales \cite{Duine2005ifi, Leblanc2009rfg, Conduit2009ipf, Jo2009ifi, Chang2011fit, Sanner2011cap} appears to be an intriguing near-future prospect.

{\bf Acknowledgements.} We thank A. Sidorov for contributions in the early stage of the experiments, and T. Enss, S. Giorgini, W. Ketterle, J. Levinsen, C. Lobo, D. Petrov, A. Recati, R. Schmidt, J. Song, C. Trefzger, P. Zoller, W. Zwerger, M. Zwierlein, and in particular M. Baranov for many stimulating discussions. We acknowledge support by the Austrian Science Fund FWF through the SFB FoQuS. M.Z.\ is supported within the Lise Meitner program of the FWF. P.M.\ is indebted to M. Lewenstein for support through the ERC Advanced Grant QUAGATUA.



\section{Methods}

{\bf Experimental conditions.}
Our system consists of $2 \times 10^4$ \K~atoms and $3.5 \times 10^5$ \Li~atoms confined in an optical dipole trap. The trap is realized with two crossed beams derived from a 1064\,nm single-mode laser source. The measured trap frequencies for Li (K) are $\nu_r = 690$\,Hz (425\,Hz) radially and $\nu_z = 86$\,Hz (52\,Hz) axially; this corresponds to a cigar-shaped sample with an aspect ratio of about 8. The preparation procedure is described in detail in Ref.~\cite{Spiegelhalder2010aop}. The Fermi energies, according to the common definition $E_F = h \sqrt[3]{6 N \nu_r^2 \nu_z}$ for harmonic traps, are $E_F^{\rm Li} = h \times 44\,{\rm kHz} = k_B \times 2.1\,\mu{\rm K}$ and $E_F^{\rm K} = h \times 10.4\,{\rm kHz} = k_B \times 500\,{\rm nK}$. At a temperature $T \approx 290$\,nK the \Li~component forms a deeply degenerate Fermi sea ($k_B T/E_F^{\rm Li} \approx 0.14$) while the \K~component is moderately degenerate ($k_B T/E_F^{\rm K} \approx 0.6$).

{\bf Effective Fermi energy.}
The \K atoms experience a nearly homogeneous \Li environment. This is because the optical trapping potential for \K~is about $2$ times deeper than for \Li and the \K~cloud is confined in the center of the much larger \Li Fermi sea \cite{Trenkwalder2011heo}. This allows us to describe the system in terms of the {effective Fermi energy} $\epsilon_F$, defined as the mean Fermi energy experienced by the \K atoms. We find $\epsilon_F = h \times 37\,$kHz, with two effects contributing to the fact that this value is about 15\% below $E_F^{\rm Li}$. The finite temperature reduces the Li density in the trap center, leading to a peak local Fermi energy of $h\times40$\,kHz. Moreover, the \K atoms sample a small region around the trap center, where the density and local Fermi energy are somewhat lower. The distribution of Fermi energies experienced by the \K cloud, i.e.\ the residual inhomogeneity of our system, can be quantified by a standard deviation of $h\times 1.9$\,kHz.

{\bf Concentration.}
The mean impurity concentration (mean density ratio $n_{\rm K}/n_{\rm Li}$) is about $0.4$, if one considers the population of K atoms in both spin states. This may be {\em a priori} too large to justify the interpretation of our data in terms of the low-concentration limit. We find that this interpretation is nevertheless valid, as we take advantage of several facts. Under strongly interacting conditions only a fraction of the K atoms is transferred into spin state $|1\rangle$ (see Fig.~2), which reduces the concentration of interacting impurities.
A recent quantum Monte Carlo calculation of the equation of state of a zero-temperature \Li-\K Fermi-Fermi mixture \cite{Gezerlis2009hlf} further supports our interpretation in the low-concentration limit: The strongest interaction in the mass-imbalanced system is expected when one has about 4 times more \K atoms than \Li atoms, and for concentrations up to a value of 1 the interaction energy per \K atom is expected to remain essentially constant. To support our basic assumption with experimental data, we also took rf spectra for variable numbers of \K atoms, confirming that in the relevant parameter range finite concentration effects remained negligibly small.

{\bf Interaction control via Feshbach resonance.}
The Feshbach resonance used for interaction tuning is discussed in detail in Refs.~\cite{Naik2011fri, Trenkwalder2011heo}. It is present for \Li in the lowest spin state and \K in the third-to-lowest spin state. The latter represents our interacting state $|1\rangle$; the corresponding quantum numbers are $F=9/2$ for the hyperfine and $m_F=-5/2$ for the magnetic sub-state. The neighboring state with $m_F=-7/2$ serves as state $|0\rangle$; here the interspecies scattering length (about $+65\,a_0$ with $a_0$ being Bohr's radius) is so small that it can be neglected to a good approximation. The tunable scattering length for state $|1\rangle$ in the Fermi sea is well described by the standard formula $a = a_{\rm bg} (1-\Delta B/(B-B_0))$ with $a_{\rm bg} = 63.0\,a_0$, $\Delta B = 880\,$mG, and $B_0 = 154.719(2)\,$G. Note that the value given for $B_0$ refers to the particular optical trap used in the experiments, as it includes a small shift induced by the trapping light. The value therefore somewhat deviates from the one given in Refs.~\cite{Naik2011fri, Trenkwalder2011heo}. In free space, without the light shift, the resonance center is located at 154.698(5)\,G.
The character of the resonance is closed-channel dominated \cite{Chin2010fri}. Following the definition \cite{Petrov2004tbp} of a range parameter $R^* = \hbar^2/(2 m_r a_{\rm bg} \, \delta \mu \, \Delta B)$, with $m_r = m_{\rm Li}m_{\rm K}/(m_{\rm Li} +  m_{\rm K})$ being the reduced mass and $\delta \mu$ the differential magnetic moment, the resonance is characterized by $R^* = 2700\,a_0$. This value accidentally lies very close to $1/\kappa_F = 2850\,a_0$, which also means that the strongly interacting regime roughly corresponds to the universal range of the resonance. Our system therefore represents an intermediate case ($\kappa_F R^* = 0.95$), where the behavior is near universal, but with significant effects arising from closed-channel contributions.

{\bf Details on Rf pulses.}
For taking the data of Fig.~2 we used Blackman pulses \cite{Kasevich1992lcb} to avoid side lobes in the spectrum. For the upper panel, the pulses were 1\,ms long (spectral width $0.7\,{\rm kHz} \simeq 0.02\,\epsilon_F/h$) and the rf power was adjusted such that $\pi$-pulses would be realized in the absence of interactions with the Fermi sea. For the data in the lower panel, the rf power was increased by a factor of 100 and the pulse duration was set to 0.5\,ms. This resulted in pulses with an area of 5$\pi$ without the Fermi sea.
For the lifetime measurements in Fig.~3, we used a sequence of 3 Blackman pulses. The first pulse (duration between $150\,\mu$s and $500\,\mu$s) was set to drive the non-interacting impurity from spin state $|0\rangle$ ($m_F=-7/2$) into state $|1\rangle$ ($m_F=-5/2$); here the frequency was carefully set to resonantly create repulsive polarons and the pulse area was set to fulfill the $\pi$-pulse condition. The second pulse was a short ($60\,\mu$s) cleaning pulse, which removed the population remaining in $|0\rangle$ by transfer to another, empty spin state ($m_F=-9/2$). The third pulse had the same parameters as the first one and resonantly back-transferred the population from the polaronic state in $|1\rangle$ to the non-interacting state $|0\rangle$, where it was finally measured by spin-state selective absorption imaging.
The measurements of Rabi oscillations in Fig.~4 were performed with simple square pulses.


\section{Supplementary Information}

\subsection{1. Theoretical framework}
\label{theoryIntro}

The theoretical results presented in the main text and in this Supplementary Information are obtained from a model that describes the behaviour of a single impurity embedded in a Fermi sea with tuneable $s$-wave interaction near a Feshbach resonance with arbitrary effective range. Two different wavefunctions are needed, depending on whether one is interested in the polaron \cite{Chevy2006upd,Combescot2007nso} or molecule \cite{Mora2009gso,Punk2009ptm,Combescot2009ato} properties. The quasiparticle parameters
for the polaron (energy $E_+$ and $E_-$, residue $Z$, effective mass) and the molecule properties can be found either variationally, or diagrammatically using the ladder approximation. Both approaches yield identical results, which closely match independent Monte-Carlo calculations \cite{Prokofev2008fpp}. The properties of the repulsive polaron, which is intrinsically unstable due to the presence of the molecule-hole continuum (MHC) and of the attractive polaron, are obtained from the self energy. In particular, the interaction induced energy shift and the decay rate are given by the real part and twice the imaginary part of the self energy, respectively \cite{Massignan2011rpa}.

Previous treatments \cite{Chevy2006upd,Combescot2007nso,Mora2009gso,Punk2009ptm,Combescot2009ato,Prokofev2008fpp,Massignan2011rpa} were based on a universal scattering amplitude, describing broad Feshbach resonances. To include effects of the finite effective range we employ a many-body T-matrix given by \cite{Bruun2005msa, Massignan2008esn}
\begin{equation}
T({\bf K},\omega)=\left[\frac{m_{r}}{2 \pi \hbar^2 \tilde{a}({\bf K},\omega)}-\Pi({\bf K},\omega)\right]^{-1}.
\label{T_mu_abg}
\end{equation}
Here $\hbar {\bf K}={\bf p}_{\rm K}+{\bf p}_{\rm Li}$ is the total momentum with ${\bf p}_{\rm Li}$ and ${\bf p}_{\rm K}$ the momenta of \Li and \K, $m_r=m_{\rm Li} m_{\rm K}/(m_{\rm Li}+m_{\rm K})$ the reduced mass, $\Pi(K,\omega)$ the \Li-\K pair propagator in the presence of the Fermi sea, and $\tilde a({\bf K},\omega)\equiv a_{\rm bg}\left(1-\frac{\Delta B}{B-B_0-E_{\rm CM}/\delta\mu}\right)$ an energy-dependent length parameter, with $a_{\rm bg}$, $\Delta B$, $B_0$, and $\delta\mu$ being the background scattering length, the width, the center, and the relative magnetic moment of the Feshbach resonance. $E_{\rm CM}({\bf K},\omega)=\hbar\omega-\hbar^2{\bf K}^2/(2M)+\epsilon_F$, with $M=m_{\rm Li}+m_{\rm K}$, is the energy in the center of mass reference frame of the colliding pair. In vacuum and close to resonance, the scattering amplitude of our model has the usual low energy expansion
\begin{equation}
-f_k^{-1}=a^{-1}+ik-r_{e} k^2/2+\ldots,
\label{equ_expansion}
\end{equation}
with the relative momentum $\hbar {\bf k}=(m_{\rm Li}{\bf p}_{\rm K}-m_{\rm K}{\bf p}_{\rm Li})/M$. The effective range is approximated by $r_{e}\approx-2R^*(1-a_{\rm bg}/a)^2$, where we introduce the range parameter $R^\ast=\hbar^2/(2m_r a_{\rm bg}\Delta B \delta\mu)$, see Ref.~\cite{Petrov2004tbp}. A detailed theoretical analysis of this model will be given elsewhere \cite{Massignan2011ppp}.

\subsection{2. Polaron peak in the spectral response}
\label{sec_peak}

\begin{figure} [!h]
\begin{center}
\hspace{+75pt}\includegraphics[width=8cm]{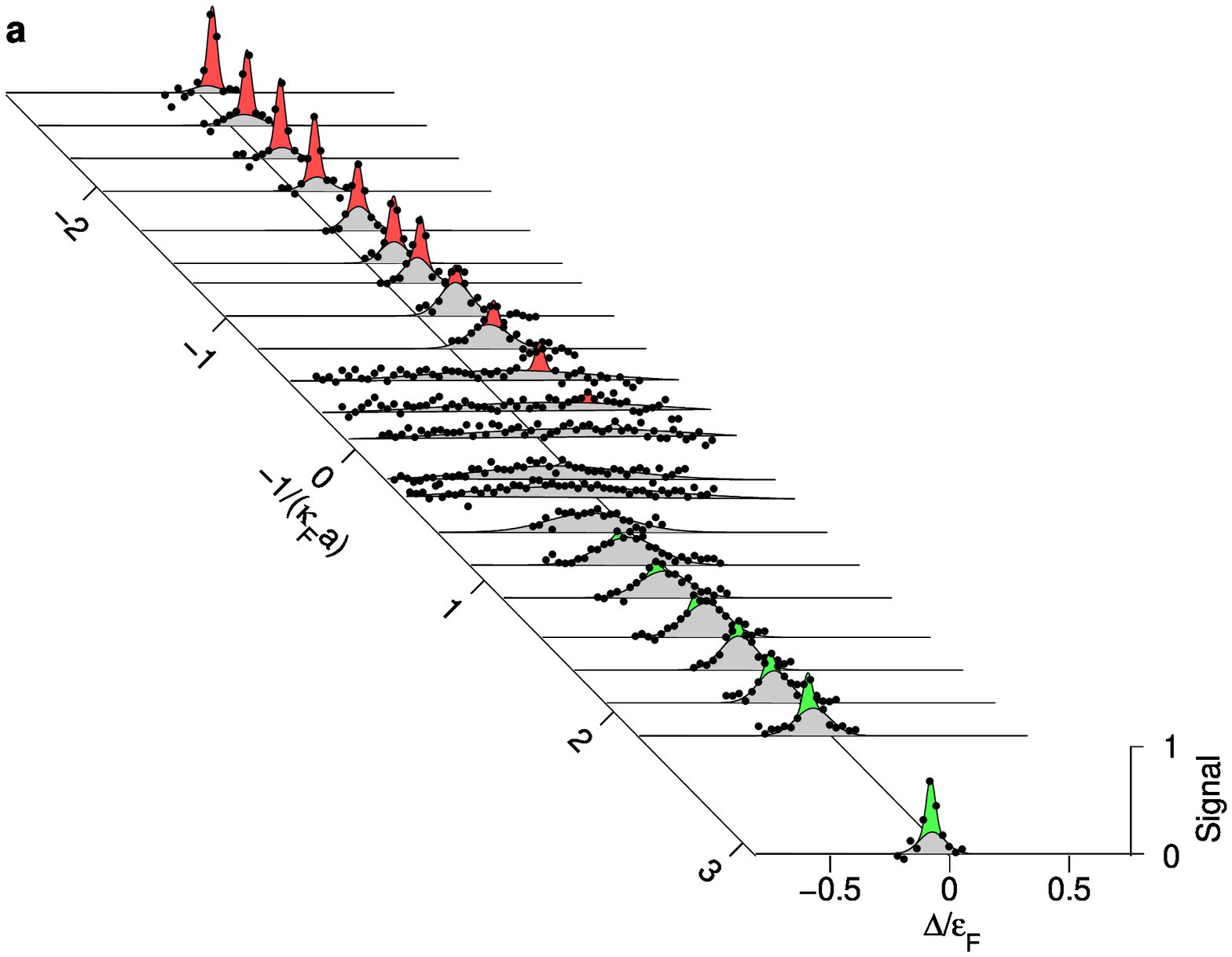}
\includegraphics[width=7cm]{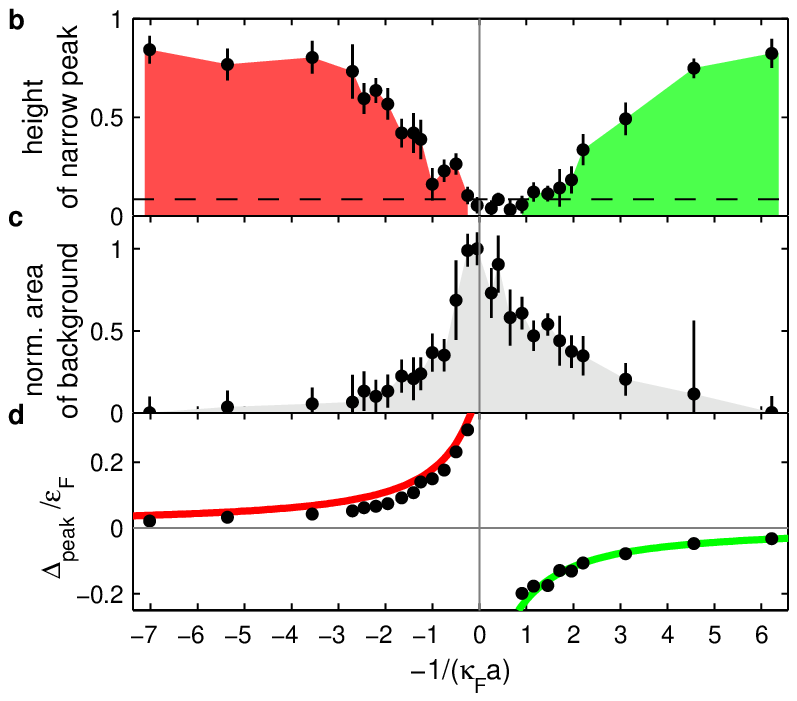}
\caption{Double Gauss analysis of the low-power spectra. The data are the same as presented in Fig.~2a plus additional data in an extended range of $1/(\kappa_F a)$. (a) The Gauss function fitting the wide background is shaded grey. The Fourier-limited Gauss function, fitting the narrow peak, is coloured red (green) along the repulsive (attractive) polaron branch. We identify the narrow peak with one of the polaron branches only if its maximum signal exceeds a threshold value of 0.085, corresponding to two times the standard deviation of the noise in our data. Any smaller peak may be caused by fitting to a noise component. The lower panels show (b) the maximum signal of the narrow peak with the dashed line indicating the threshold, (c) the area under the wide Gauss function normalized to its maximum value, and (d) the detuning at the center of the narrow peak, provided that the peak signal exceeds the threshold, compared to the theoretical calculation of $E_+$ and $E_-$ (red and green line). The error bars indicate the fit uncertainties.} \label{fig_peak} \end{center} \end{figure}

The spectra in Fig.~2a of the main text show a narrow, {\it coherent} peak on top of a spectrally broad, {\it incoherent} background. Here, we investigate these two spectral parts in more detail. Note that the background is actually better visible in Fig.~2b, but these spectra do not allow for a quantitative comparison of the two parts because of the strong saturation of the narrow polaron peaks.

The narrow peak stems from the attractive or repulsive polarons, which correspond to well defined energy levels, provided that the lifetime of the quasiparticle exceeds the pulse duration. As a consequence, the lineshape is expected to be Fourier limited except for the rapidly decaying repulsive polarons very close to resonance. In contrast, the background is spectrally wide, on the order of $\varepsilon_F$. The main contribution to the background stems from the MHC. Another contribution may arise from the excitation of additional particle-hole pairs in the Fermi sea when transferring to a quasiparticle with a momentum that is different from the momentum of the impurity in the initial state.

We distinguish between the narrow peak and the wide background by means of a double Gauss fit. Vertical cuts through Fig.~2a are presented in Fig.~\ref{fig_peak}a together with the fit curves. The width $\sigma_p$ of the Gauss function fitting the narrow peak is fixed to the one associated with the Gaussian fit of the Blackman pulse line shape used in the experiment, $\sigma_p=0.7\,$kHz$=0.019\,\varepsilon_F/h$. We constrain the width $\sigma_b$ of the Gauss function reproducing the background to $3\times0.019\,\varepsilon_F/h<\sigma_b<0.5\,\varepsilon_F/h$. The lower bound avoids the misinterpretation of the narrow peak as background and the upper bound, corresponding to the maximal width of the continuum as obtained from the spectra in Fig.~2b, avoids unphysically large values of $\sigma_b$ when the background signal is weak. We find that the narrow peak dominates for weak positive and negative interaction strength while the wide background dominates in the strongly interacting regime. This trend is shown in Fig.~\ref{fig_peak}b and Fig.~\ref{fig_peak}c, where we present the maximum signal of the narrow peak and the area of the background, respectively. Note, that the signal in Fig.~\ref{fig_peak}b is proportional to the area of the narrow peak since $\sigma_p$ is kept constant. Figure \ref{fig_peak}d shows the detuning at the center of the narrow peak, which corresponds to the energy of the quasiparticles. The measured energies agree remarkably well with the calculation. The slight mismatch between theory and experiment may be attributed to systematic errors in the determination of $\varepsilon_F$ and $B_0$.

The area of the wide background exhibits a maximum close to $-1/(\kappa_Fa)=\,0$, but it shows an asymmetry as it falls off significantly slower on the attractive ($a<0$) side, see Fig.~\ref{fig_peak}c. We attribute this asymmetry to the narrow character of the Feshbach resonance. The interaction becomes resonant when the real part of the inverse scattering amplitude, given in Eq.~\ref{equ_expansion}, is zero. This leads to the resonance condition $a_{\rm res}^{-1}=r_{e} k^2/2$, where $a_{\rm res}$ is the value of the scattering length at which the interaction becomes resonant. In the limit of a broad resonance with $r_{e}=0$, this condition is fulfilled for any $k$ at the center of the resonance, where the scattering length diverges. However, at a narrow resonance with $r_{e}<0$ the condition requires a negative $a_{\rm res}$ for $k>0$. The mean square momentum in the Fermi sea is $3/5\times\kappa_F^2$, leading to a mean square relative momentum of $3/5\times(40/46\times\kappa_F)^2$. Using this value for $k^2$, and inserting $r_e\approx-2R^\ast$
in the above resonance condition, we obtain $-1/(\kappa_Fa)=\,0.43$. This represents an effective shift of the Feshbach resonance center, as we average over all momenta of the Fermi sea \cite{Ho2011aro}. The magnitude of this shift agrees well with the observed asymmetry. Moreover, we find that many features at our narrow resonance appear to be shifted, e.g.~the polaron-to-molecule crossing. However, the narrowness has many more implications and cannot simply be reduced to this shift. We will come back to this point in the context of the lifetime of the repulsive polaron, see Sec.~4.

The repulsive polaron peak is clearly visible up to $-1/(\kappa_F a)\approx-0.3$ while the attractive polaron peak vanishes already at $-1/(\kappa_F a)\approx0.9$, see Fig.~\ref{fig_peak}b. The fading out of the quasiparticle peak towards the strongly interacting regime approximately coincides with the position where the quasiparticle branches merge into the MHC. This shows that the polaron state is hardly observable as soon as it becomes degenerate with molecule-hole excitations. The MHC is not strictly limited to the range from $E_m$ to $E_m-\varepsilon_F$, as discussed in more detail in Sec.~3. It extends below $E_m-\varepsilon_F$ because of finite temperature effects. It also extends slightly above $E_m$ because of additional excitations in the spectral function of the molecules \cite{Schmidt2011esa}. As a consequence, for finite temperature, the attractive polaron can become degenerate with molecule-hole excitations for values of the interaction parameter above the calculated polaron-to-molecule crossing. This explains that the observed sharp peak is observed to disappear already at $-1/(\kappa_Fa)\approx0.9$, which lies somewhat above the zero-temperature polaron-to-molecule crossing predicted at $0.6$.

It is interesting to consider the data analysis presented in Fig.~\ref{fig_peak}b and c in relation to the common method of extracting the quasiparticle residue $Z$ from the spectral weight of the narrow peak \cite{Ding2001cqw}. Close to resonance, we are in the linear response regime and our data can be interpreted in terms of this method. Our data suggests that this method leads to a significant underestimation of $Z$. For example at $-1/(\kappa_F a)\approx0.9$, where the narrow peak of the attractive polaron vanishes, our theory still predicts $Z\approx0.7$. This underestimation is consistent with the one reported in Ref.~\cite{Schirotzek2009oof}, see also related discussion in Ref.~\cite{Punk2009ptm}. A plausible explanation may be that such a method does not probe the polaron states alone, but also the molecule-hole excitations, which are degenerate with the polaron state. Our alternative method of measuring the residue via the Rabi frequency, as presented in the main paper, offers the advantage of being much less affected by the molecule-hole contribution. In fact, only the coherent part of the quasiparticle is expected to produce Rabi oscillations, see Sec.~6.

\subsection{3. Molecule-hole continuum}
\label{sec_MHC}

\begin{figure} [h]
\begin{center}
\includegraphics[width=8cm]{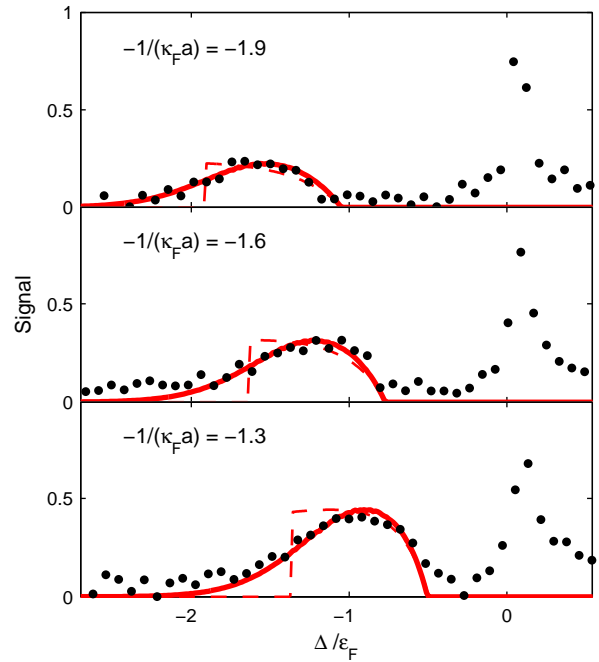}
\caption{Molecule association spectra for different values of the interaction parameter. The signal is the fraction of transferred atoms as a  function of the rf detuning. The data correspond to vertical cuts through Fig.~2b. The dashed line is the line shape model for zero temperature and the solid line for finite temperature. The upper threshold of the theoretical spectra corresponds to $E_m$.}
\label{fig_MHC} \end{center} \end{figure}

The spectra presented in Fig.~2b of the main text reveal the MHC. This continuum arises from processes where the rf field associates a \K impurity and a \Li atom out of the Fermi sea to a molecule. Here we present a simple model for the spectral line shape, which allows us to interpret the data up to $-1/(\kappa_F a)\approx-1$, see Fig.~\ref{fig_MHC}.

For modeling the line shape, we consider two-body processes in which the rf field associates one \K and one \Li atom to a molecule. Higher-order processes, involving more than two particles, are neglected in this model but are briefly discussed at the end of this section.  Let us first consider the association of \Li and \K with momenta $p_{\rm Li}=p_{\rm K}=0$. This results in a molecule at rest plus a Fermi sea with a hole in the center. The energy of this state is determined by the binding energy of the molecule and by the interaction of the molecule with the Fermi sea. It is given by $E_m$ and sets the onset of the MHC from the right (the top) in Fig.~\ref{fig_MHC} (Fig.~2b). In general, \Li and \K have finite initial relative momentum $\hbar k$, leading to an initial relative kinetic energy in the center of mass frame $E_r=\hbar^2k^2/2 m_r$. The energy conservation of the association process is expressed in the Dirac $\delta$ function in Eq.~\ref{equ_mhc}. As a consequence, the molecule spectrum extends downwards to energies below $E_m$. We now consider an ensemble of \K and \Li atoms. Our experimental conditions are well approximated by a thermal cloud of \K in a homogeneous Fermi sea of \Li (see Methods). The momentum distribution of \Li is given by the Fermi-Dirac distribution $f^{\rm FD}_{\rm Li}(E_{\rm Li})$, with $E_{\rm Li}=p_{\rm Li}^2/2m_{\rm Li}$. The one of \K is approximated by the Maxwell-Boltzmann distribution $f^{\rm MB}_{\rm K}(E_{\rm K})$, with $E_{\rm K}=p_{\rm K}^2/2m_{\rm K}$. The latter distribution does not change its momentum dependence with position, thus, no integration over space is needed to obtain the spectral response
\begin{equation}
\label{equ_mhc}
\begin{split}
\mathcal{S}(\Delta)\propto \int\int d^3p_{\rm Li} ~~d^3p_{\rm K}
~~f^{\rm FD}_{\rm Li}(E_{\rm Li}) \\ ~~f^{\rm MB}_{\rm K}(E_{\rm K}) ~~\mathcal{F}(k) ~~\delta(-E_m+E_r+\Delta),
\end{split}
\end{equation}
where $\mathcal{F}(k)$ is the Franck Condon overlap of the initial wavefunction with the molecule wavefunction. In our case the interaction in the initial state is negligible and $\mathcal{F}(k)$, as given in Ref.~\cite{Chin2005rft}, reduces to $\mathcal{F}(k)\propto (E_r/E_b^3)^{1/2}(1+E_r/E_b)^{-2}$. The parameter $E_b$ is the binding energy of a molecule in vacuum at a resonance with finite effective range and reads $E_b=\hbar^2/(2 m_r a^{\ast2}) $ with the parameter \cite{Petrov2004tbp} $a^\ast=-r_e/(\sqrt{1-2r_{e}/a}-1)$. In the calculation of $\mathcal{F}(k)$, we do not account for interactions with the Fermi sea. Because of this approximation, we apply the model only for $-1/(\kappa_F a)<-1$. For fitting the model line shapes to the experimental data, adjustable parameters are the individual heights of the spectra and the center of the Feshbach resonance. The latter parameter is required to be the same for all data sets in Fig.~\ref{fig_MHC}. Independently determined parameters are $k_B T/\varepsilon_F=0.16$ and $\varepsilon_F=h\times 37$\,kHz. The model (solid lines) reproduces our data remarkably well. It allows us to pinpoint the resonance position to $B_0 = 154.719(2)\,$G. This determination of $B_0$ relies on our theoretical model to calculate $E_m$. To test this model dependence, we replace $E_m$ simply by the binding energy of the molecule in vacuum plus the mean field energy, considering the corresponding atom-dimer scattering length \cite{Levinsen2011ada}. Using this simple model, the fit yields a resonance position that is 1\,mG higher, which shows that the model dependence causes only a small systematic uncertainty. Moreover, the statistical fit uncertainty and the field calibration uncertainty are about 1\,mG each.

For $T=0$ and all other parameters unchanged, the model provides the dashed lines in Fig.~\ref{fig_MHC}. The spectra show a sharp drop at $\Delta=E_m-(40/46)\,\varepsilon_F$, which corresponds to the association of an impurity at rest and a majority atom at the Fermi edge.
In an equal-mass mixture this process would occur at $\Delta=E_m-(1/2)\,\varepsilon_F$. Thus, the width of the MHC in the two-body approximation is much larger for a heavy impurity than it is for an equal-mass impurity and it is even narrower for a light impurity.

The true zero temperature ground state is actually at the energy $E_m-\varepsilon_F$, a molecule at rest formed from a \K atom at rest and a \Li atom at the Fermi edge. However, to reach this state, momentum conservation requires a higher-order process, i.e.~the scattering of at least one additional \Li atom from and to the Fermi surface. Such processes are not included in the model presented here, which only considers the direct association of two atoms by an rf photon.

In the strongly interacting regime the spectral function of the molecule shows additional excitations above the molecular ground state \cite{Schmidt2011esa}. This leads to an extension of the MHC spectral response above $E_m$, of which we find clear indications in our data. The lower panel in Fig.~\ref{fig_MHC} shows finite signal above $E_m$ and the extension above $E_m$ is very evident in the strongly interacting regime, see Fig.~2b.

\subsection{4. Decay rate of the repulsive polarons}
\label{sec_decayrate}

We analyse the decay of the repulsive polarons by assuming that they decay into well defined attractive polarons or well-defined molecules.
In this  quasiparticle picture, the decay is associated with the formation of a particle-hole pair in the Fermi sea to take up the released energy. In this sense,
the decay into the attractive polaron is a 2-body process and the decay into the molecule is a 3-body process. We calculate the decay rate for these two channels by including them into the polaron self energy using a pole expansion of the \K propagator writing $G({\mathbf k},\omega)\simeq Z_+/(\hbar\omega-E_+-\hbar^2k^2/(2m_{\rm K}))+Z_-/(\hbar\omega-E_--\hbar^2k^2/(2m_{\rm K}))$ and
a pole expansion of the T-matrix writing $T({\mathbf k},\omega)\simeq Z_mg^2/(\hbar\omega-(E_m-\varepsilon_F)-\hbar^2k^2/(2M))$. Here,  $Z_{\pm}$ is the quasiparticle residue of the repulsive and attractive polaron respectively and $Z_m$ the quasiparticle residue of the molecule. The factor $g^2=2\pi\hbar^4 /(m_r^2a^\ast\sqrt{1-2r_{e}/a})$ is the residue of the vacuum T-matrix for a general resonance. The details of this approach are given in Refs.~\cite{Massignan2011rpa,Bruun2010dop}, the only difference being that here we include the effects of the finite effective range. The imaginary part of the self energy gives the decay rate of the wavefunction and we thus take twice the imaginary part to calculate the population decay. The 2-body decay into the attractive polaron and an additional particle-hole pair
is calculated numerically to all orders in the T-matrix by inserting the pole expansion for the \K propagator in the self energy in the ladder approximation. For the 3-body decay  into a molecule and an additional particle-hole pair, we include terms containing two \Li holes in the \K self energy \cite{Bruun2010dop}, and an expansion to second order in the T-matrix relevant for $-1/(\kappa_Fa)\ll-1$ yields
\begin{equation}
\begin{split}
\Gamma_{PM}\simeq\frac{64\kappa_Fa}{45\pi^3}\frac{Z_{+}^3}{m_{\rm K}^2\sqrt{m_{\rm Li}}}\left(1+\frac{m_{\rm Li}}{M}\right)^{3/2}\\ \left(\frac{\hbar\kappa_F}{\sqrt{2(E_+-E_m+\epsilon_F)}}\right)^{5}
\frac{a}{a^\ast\sqrt{1-2r_{e}/a^\ast}}\frac{\epsilon_F}{\hbar}.
\end{split}
\label{3bodyDecay}
\end{equation}
For simplicity, we have taken $Z_m=1$, which is an appropriate assumption for $-1/(\kappa_Fa)\ll-1$. The effect of the narrow resonance on the decay rate enters through the quasiparticle residue $Z_+$, the energies $E_+$, $E_-$, $E_m$ and directly through the effective range $r_{e}$. This decay rate has the same $a^6$ dependence as the three-body decay in vacuum in the limit of a broad resonance derived in Ref.~\cite{Petrov2003tbp}. The numerical prefactor however differs since we have included the effects of the Fermi sea in a perturbative calculation.

\begin{figure} [t]
\begin{center}
\includegraphics[width=8cm]{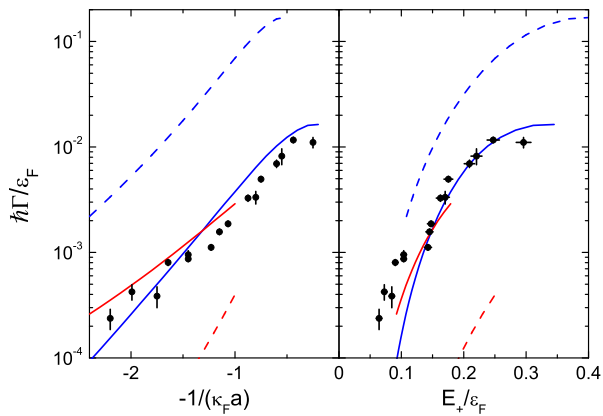}
\caption{Decay rates of repulsive \K polarons in a Fermi sea of \Li atoms, shown as a function of interaction strength (left) and of the energy of the repulsive polaron (right). Blue and red lines represent the two- and three-body contributions, respectively, while data points are the experimental findings as also shown in Fig.~3 of the main text. The results for the moderately narrow resonance under study here (solid lines) is compared with the theoretical results obtained for the universal limit of a very broad resonance (dashed lines). The experimental values of $E_+$ are obtained by interpolation of the narrow peak position data $\Delta_{\rm peak}$, see Fig.~\ref{fig_peak}d.}
\label{fig_comparedecay} \end{center} \end{figure}

The results for the decay rates of repulsive polarons are shown in Fig.~\ref{fig_comparedecay}. The experimental data agree well with the theoretical results obtained for our narrow resonance (continuous lines) as already shown in Fig.~3 in the main text. For comparison, we also show the decay rates one would obtain in the limit of a broad resonance (dashed lines). We find that as magnitude of the effective range increases with respect to the interparticle spacing, the dominant two-body decay is strongly suppressed. This suppression is mainly due to a large reduction of the attractive polaron residue $Z_{-}$. Instead, the weaker three-body decay increases, which we attribute to the reduction of the polaron-molecule energy difference $E_+-E_m+\epsilon_F$. Taking both decay rates together, the decay rate is at least an order of magnitude smaller at our narrow resonance as compared to the case of a broad resonance. It is important to note that this strong suppression of the decay at a given $-1/(\kappa_Fa)$ cannot be simply attributed to the effective resonance shift at our narrow Feshbach resonance as discussed in Sec.~2. When taking this shift into account, a suppression factor of five to ten remains. To highlight this point, we choose a representation that is independent of the interaction parameter and that gives the dependence on the polaron energy, a direct manifestation of strong interactions. The right panel shows the same data and calculations as a function of $E_+$. Also for a given $E_+$, the repulsive polaron at our narrow resonance turns out to be much more stable than the repulsive polaron at a broad resonance.

\subsection{5. Decay of repulsive polarons to molecules}

The decay of the repulsive polarons, shown in Fig.~3 of the main text, is measured by applying a special three-pulse scheme (see Methods). In this section we exploit the flexibility of this scheme to study the decay to lower-lying energy states in more detail. At a given interaction strength $-1/(\kappa_F a)=-0.9$, we demonstrate that the repulsive polarons decay to molecules by showing that an rf spectrum taken after decay perfectly matches a reference spectrum of molecules.

\begin{figure} [h]
\begin{center}
\includegraphics[width=6cm]{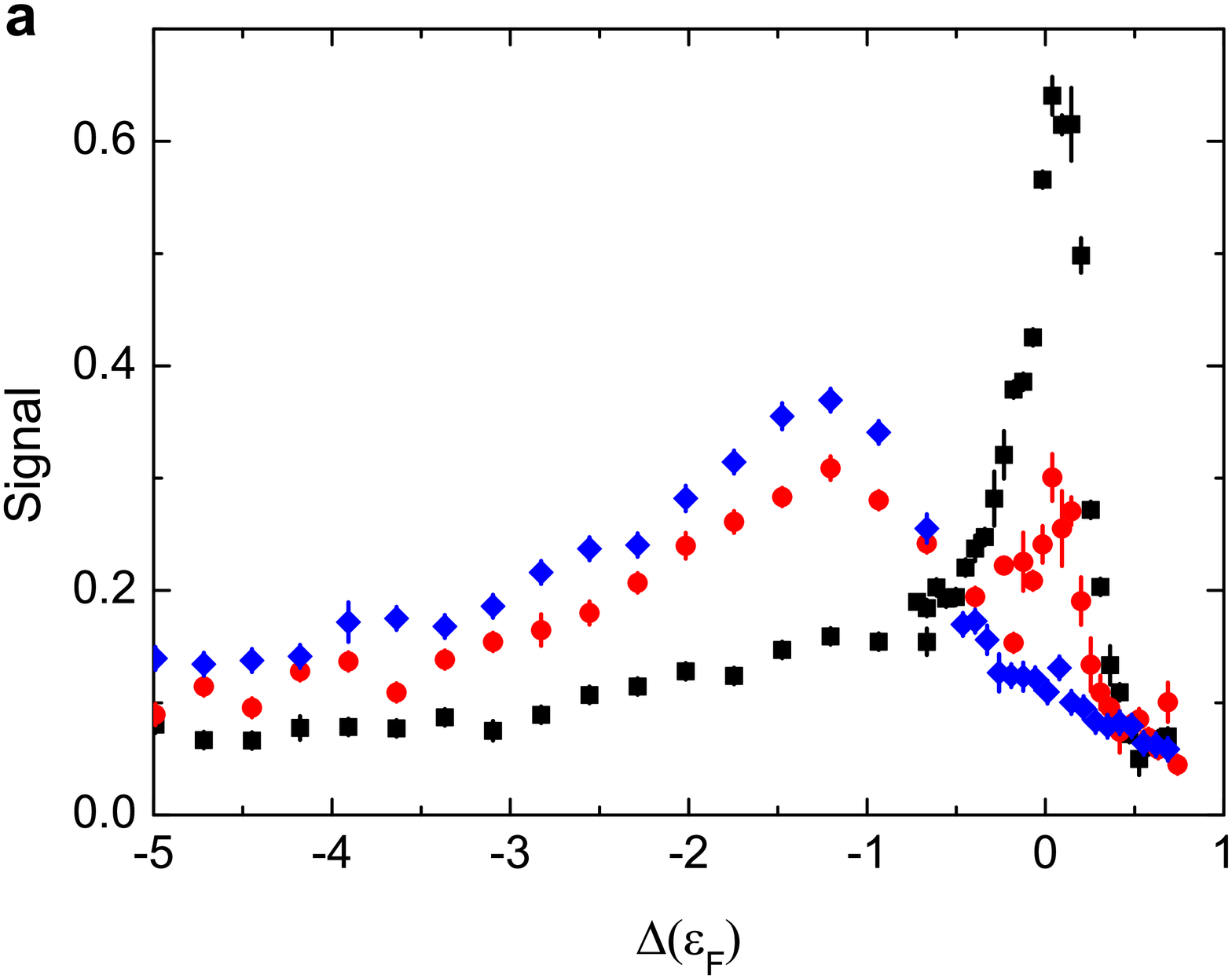}
\includegraphics[width=6cm]{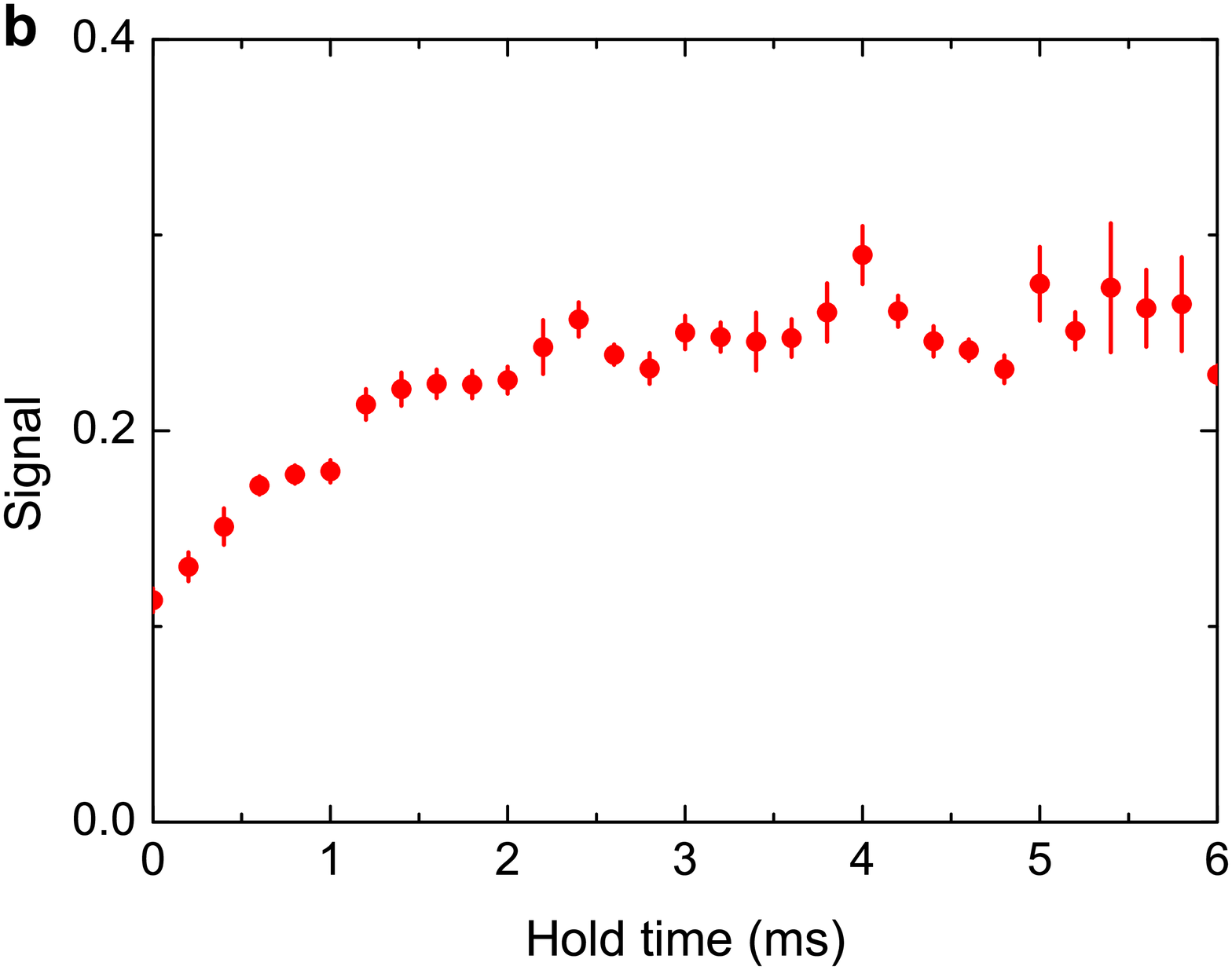}
\caption{Decay of repulsive polarons to molecules at $-1/(\kappa_F a)=-0.9$. (a) The black squares (red dots) show the spectrum right after ($2\,$ms after) the repulsive polaron has been populated. The blue diamonds show the dissociation spectrum of molecules for reference. The signal is the fraction of atoms transferred from the interacting spin state \ket{1} to the noninteracting spin state \ket{0}. Note that the polaron peak at positive detuning is highly saturated and thus its signal is not proportional to the number of polarons. (b) The rf energy detuning is fixed to $\Delta=-1.3\,\varepsilon_F$ and the signal is recorded versus hold time. The error bars indicate the statistical uncertainties derived from at least three individual measurements.}
\label{fig_decayproducts}
\end{center}
\end{figure}

To populate the repulsive polaron branch, as done for the measurements of the decay rate, we tune the energy of the first pulse to $E_+$, corresponding to $\Delta=0.16\,\varepsilon_F$ at $-1/(\kappa_F a)=-0.9$. The pulse duration ($t_p=0.06\,$ms) and the intensity are set to correspond to a $\pi$-pulse in the noninteracting system. The second pulse removes the remaining non-transferred atoms by transferring them to a third spin state. In contrast to the decay measurement presented in the main text, we here use much more rf power for the third pulse to be able to efficiently dissociate molecules. For this purpose, we set $t_p=0.3\,$ms and the pulse area corresponds to a $3\pi$-pulse in the noninteracting system. By varying the rf detuning, we record spectra for zero hold time (black squares) and for a hold time of $2\,$ms (red dots), see Fig.~\ref{fig_decayproducts}a. The peak at small positive detuning shows the back-transfer of repulsive polarons. The corresponding signal decreases with hold time, signalling the decay of the repulsive polaron. In addition, a wide continuum in a range of negative detunings rises with increasing hold time. Such a wide continuum involves coupling to high momentum states, signaling a short distance between \K and $^6$Li. To confirm that this continuum stems from molecules, we compare it to a reference spectrum of the dissociation of molecules (blue diamonds). We find a perfect match. To take such a reference spectrum, only the detuning of the first rf pulse is changed to directly associate molecules in the MHC instead of populating the repulsive polaron branch. We achieve a good association efficiency with $\Delta=-0.54\,\varepsilon_F$ and $t_p=0.5\,$ms.

To study the evolution of the molecule population, which is fed by the decay of the repulsive polarons, we set the detuning of the third pulse to the peak of the molecule signal at $\Delta=-1.3\,\varepsilon_F$ and record the signal as a function of the hold time, see Fig.~\ref{fig_decayproducts}b. A simple exponential fit yields a rate of about 1\,ms$^{-1}$=\,0.0043\,$\varepsilon_F/\hbar$, which is in good agreement with the measured decay rate of the repulsive polaron at $-1/(\kappa_F a)=-0.9$. The finite signal at zero hold time may have two origins. One contribution is some decay during the finite pulse durations of the three pulses, which are not included in the hold time. Another contribution may be the high momentum tail of the repulsive polarons as discussed in Ref.~\cite{Schirotzek2009oof}.

Note that we do not find any second sharp peak at negative detuning, which would indicate the population of the attractive polaron branch. In case the repulsive polaron decays to the attractive polaron, the absence of the attractive polaron peak implies a very rapid subsequent decay of the attractive polaron to the MHC. Such a fast decay of the attractive polaron to the MHC is consistent with the very small signal of the attractive polaron peak throughout the regime of strong interaction as discussed in Sec.~2.

Let us briefly discuss the possible role of inelastic two-body relaxation in the \Li-\K mixture, which is energetically possible as \K is not in the lowest spin state. This process was identified in Ref.~\cite{Naik2011fri} as a source of losses.
However, this relaxation is about an order of magnitude slower than the measured decay rate of the repulsive polaron and thus does not affect our measurements.

\subsection{6. Rabi oscillations and polaron quasiparticle residue}
\label{sec_Z}

For high rf power, the signal is well beyond linear response and the \K atoms  exhibit coherent Rabi oscillations between the spin states \ket{0} and \ket{1}. In this regime the oscillations are so fast, that the polaron decay plays a minor role and can be ignored to a first approximation. The Rabi frequency depends on the matrix element of the rf probe between the initial state \ket{0} and the final state \ket{1}. Since the probe is homogenous in space, it does not change the spatial part of the atomic wavefunction and it can be described by the operator \cite{Massignan2008tpi} $\hat R \propto \Omega_0\sum_{\bf q}(\hat a_{1\bf q}^\dagger\hat a_{0\bf q}+h.c.)$ where $\hat a^\dagger_{i{\bf q}}$ ($\hat a_{i{\bf q}}$) creates (annihilates) a \K atom with momentum ${\bf q}$ in spin state $i$ and $\Omega_0$ is the unperturbed Rabi frequency of the $|0\rangle$ to $|1\rangle$ transition in the non-interacting case. Considering for simplicity an impurity at rest, the initial non-interacting state is given by $|I\rangle=\hat a_{0{\bf q}=0}^\dagger|{\rm FS}\rangle$ where $|{\rm FS}\rangle$ is the \Li Fermi sea. The final polaronic state at zero momentum can be written as \cite{Chevy2006upd}
\begin{equation}
|F\rangle= \sqrt{Z} \hat a_{1{\bf q}=0}^\dagger|{\rm FS}\rangle +
\sum_{q<\hbar\kappa_F<p}\phi_{{\bf p},{\bf q}} \hat a_{1{\mathbf q}-{\mathbf p}}^\dagger
\hat b^{\dag}_{{\bf p}}\,\hat b_{{\bf q}}|{\rm FS}\rangle+\ldots
\label{ansatz}
\end{equation}
where $\hat b^\dagger_{\bf q}$ ($\hat b_{\bf q}$) creates (annihilates) a \Li atom with momentum ${\bf q}$. The second term contains a Fermi sea with at least
one particle-hole excitation and thus is orthogonal to an unperturbed Fermi sea. Therefore the  matrix element reduces to $\langle F|\hat R|I\rangle=\sqrt{Z}\,\Omega_0$ and we obtain the Rabi frequency
\begin{equation}
\Omega=\sqrt{Z}\,\Omega_0.
\end{equation}
We neglect the momentum dependence of the quasiparticle residue and do not perform a thermal average over the initial states, which we expect to be a good approximation
since $T\ll \epsilon_F/k_B$.

\begin{figure} [t]
\begin{center}
\includegraphics[width=8cm]{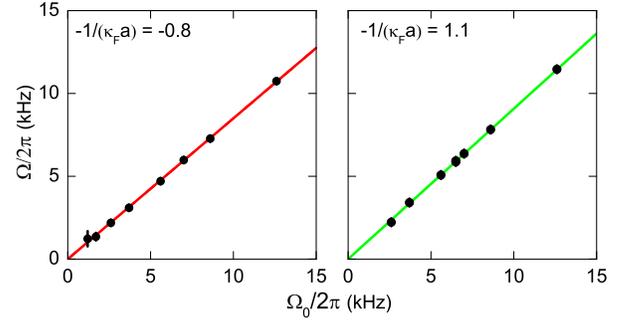}
\caption{Linear increase of the Rabi frequency $\Omega$ with the unperturbed Rabi frequency $\Omega_0$. The left (right) panel shows the driving to the repulsive (attractive) polaron. The solid lines are linear fits without offset and demonstrate the proportionality $\Omega\propto\Omega_0$.} \label{fig_omega} \end{center} \end{figure}

In Fig.~\ref{fig_omega} we plot the observed Rabi frequency $\Omega$ as a function of the unperturbed Rabi frequency $\Omega_0$. We find that the proportionality $\Omega\propto \Omega_0$ holds over a wide range of rf power. The measurements presented in the main text, taken at $\Omega_0=2\pi\times 6.5\,$kHz and 12.6\,kHz, are safely within this range.


\bibliographystyle{apsrev}




%
%



\end{document}